\documentclass[preprint]{aastex}

\newcommand{\lya}{Ly$\alpha$}
\newcommand{\nv}{N\,{\sc v}}
\newcommand{\siii}{Si\,{\sc ii}}
\newcommand{\oi}{O\,{\sc i}}
\newcommand{\civ}{C\,{\sc iv}}
\newcommand{\mgii}{Mg\,{\sc ii}}

\begin{document}

\title{\boldmath A SURVEY OF $z\sim6$ QUASARS IN THE SDSS DEEP STRIPE. II.
DISCOVERY OF SIX QUASARS AT $z_{\rm AB}>21$}

\author{Linhua Jiang\altaffilmark{1,2}, Xiaohui Fan\altaffilmark{1,3},
  Fuyan Bian\altaffilmark{1,2}, James Annis\altaffilmark{4},
  Kuenley Chiu\altaffilmark{5}, Sebastian Jester\altaffilmark{3}, 
  Huan Lin\altaffilmark{4}, Robert H. Lupton\altaffilmark{6}, 
  Gordon T. Richards\altaffilmark{7}, Michael A. Strauss\altaffilmark{6},
  Viktor Malanushenko\altaffilmark{8}, Elena Malanushenko\altaffilmark{8},
  and Donald P. Schneider\altaffilmark{9}}
\altaffiltext{1}{Steward Observatory, University of Arizona,
  933 North Cherry Avenue, Tucson, AZ 85721}
\altaffiltext{2}{Visiting Astronomer, Kitt Peak National Observatory, National 
  Optical Astronomy Observatory, which is operated by the Association of 
  Universities for Research in Astronomy (AURA) under cooperative agreement 
  with the National Science Foundation.}
\altaffiltext{3}{Max-Planck-Institut f\"{u}r Astronomie, K\"{o}nigstuhl 17,
  D-69117 Heidelberg, Germany}
\altaffiltext{4}{Fermi National Accelerator Laboratory, P.O. Box 500,
  Batavia, IL 60510}
\altaffiltext{5}{Department of Astronomy, California Institute of Technology, 
  MS 105-24, Pasadena, CA 91125}
\altaffiltext{6}{Department of Astrophysical Sciences, Princeton University,
  Princeton, NJ 08544}
\altaffiltext{7}{Department of Physics, Drexel University, 3141 Chestnut
  Street, Philadelphia, PA 19104}
\altaffiltext{8}{Apache Point Observatory, B.P Box 59, Sunspot, NM 88359}
\altaffiltext{9}{Department of Astronomy and Astrophysics, Pennsylvania
  State University, 525 Davey Laboratory, University Park, PA 16802}

\begin{abstract}

We present the discovery of six new quasars at $z\sim6$ selected from the
Sloan Digital Sky Survey (SDSS) southern survey, a deep imaging survey 
obtained by repeatedly scanning a stripe along the celestial equator. The six 
quasars are about two magnitudes fainter than the luminous $z\sim6$ quasars 
found in the SDSS main survey and one magnitude fainter than the quasars 
reported in Paper I \citep{jia08}. Four of them comprise a complete 
flux-limited sample at $21<z_{AB}<21.8$ over an effective area of 195 deg$^2$. 
The other two quasars are fainter than $z_{AB}=22$ and are not part of the
complete sample. The quasar luminosity function at $z\sim6$ is well described 
as a single power law $\Phi(L_{1450})\propto L_{1450}^{\beta}$ over the 
luminosity range $-28<M_{1450}<-25$. The best-fitting slope $\beta$ varies 
from --2.6 to --3.1, depending on the quasar samples used, with a statistical 
error of 0.3--0.4.
About 40\% of the quasars discovered in the SDSS southern survey have 
very narrow \lya\ emission lines, which may indicate small black hole masses 
and high Eddington luminosity ratios, and therefore short black hole growth 
time scales for these faint quasars at early epochs.

\end{abstract}


\section{INTRODUCTION}

High-redshift quasars serve as cosmological probes for studying the early 
universe. In recent years, about 40 quasars at $z\sim6$ have been discovered;
the most distant ones are at $z\sim6.4$ \citep{fan06,wil07}. They harbor 
billion-solar-mass black holes, and thus are essential in understanding black
hole accretion and galaxy formation in the first billion years of cosmic time. 
Most of the known quasars at $z\sim6$ were discovered from $\sim8000$ deg$^2$ 
of imaging data of the Sloan Digital Sky Survey \citep[SDSS;][]{yor00}. These 
luminous quasars ($z_{AB}\leq20$, $M_{1450}\sim-27$) were selected as 
$i$-dropout objects using optical colors; follow-up near-infrared (NIR) 
photometry and optical spectroscopy were used to distinguish against late-type 
dwarfs \citep[e.g.][]{fan01}. Several other high-redshift quasars have also 
been discovered based on their infrared or radio emission 
\citep[e.g.][]{coo06,mcg06}.

Currently ongoing surveys of $z\sim6$ quasars include the UKIRT Infrared Deep
Sky Survey \citep[UKIDSS;][]{war07} and the Canada-France High-redshift Quasar
Survey \citep[CFHQS;][]{wil05}. The UKIDSS survey is being carried out in the
$YJHK$ bands using the Wide Field Camera \citep{cas07} on the 3.8 m UKIRT 
telescope. It will survey 7500 square degrees of the northern sky. The 
resulting NIR data together with the SDSS multicolor optical data can be used 
to select high-redshift quasar candidates. The UKIDSS team has found two 
quasars at $z\sim6$ \citep{ven07,mor08}. These quasars are fainter than 
$z_{AB}=20$, the magnitude selection limit that the SDSS team used 
\citep{fan06}. The CFHQS survey is an optical imaging survey of $\sim550$ 
square degrees in the $griz$ bands on the Canada-France-Hawaii Telescope
\citep{wil09}. The survey is about two magnitudes deeper than the SDSS wide 
survey. The CFHQS team has found 10 quasars at $z\sim6$ to date, 
including the most distant quasar known at $z=6.43$ \citep{wil07,wil09}.

This paper is the second in a series presenting $z\sim6$ quasars selected from 
the SDSS southern survey, a deep imaging survey obtained by repeatedly 
scanning a 300 deg$^2$ stripe along the celestial equator. 
In \citet[hereafter Paper I]{jia08} we reported the discovery of five quasars
at $5.85\le z\le 6.12$ in 260 deg$^2$ of the deep stripe. Together with 
another quasar known in this region \citep{fan04}, they constructed a complete
flux-limited quasar sample at $20<z_{AB}<21$. Based on the combination of this 
sample, the luminous quasar sample from $\sim8000$ deg$^2$ of SDSS, and the 
\citet{coo06} sample, we found a steep slope at the bright end of the quasar 
luminosity function (QLF) at $z\sim6$. In this paper we present the discovery 
of six new quasars in the SDSS deep stripe. All six quasars are about one
magnitude fainter than those reported in Paper I, or two magnitudes fainter
than the luminous SDSS quasars. With these new quasars, we can measure the 
QLF over three magnitudes. The basic procedures of candidate
selection, follow-up observations, and data reduction are similar to the
procedures used in Paper I.

The structure of the paper is as follows. Section 2 briefly introduces the 
construction of the SDSS co-added imaging data. Section 3 describes our
selection criteria and follow-up observations of quasar candidates. 
We present the six new quasars in Section 4 and discuss the QLF at 
$z\sim6$ in Section 5. We summarize the paper in Section 6.
Throughout the paper we use a $\Lambda$-dominated flat
cosmology with H$_0=70$ km s$^{-1}$ Mpc$^{-1}$, $\Omega_{m}=0.3$, and
$\Omega_{\Lambda}=0.7$ \citep{spe07}.

\section{CONSTRUCTION OF THE SDSS CO-ADDED IMAGING DATA}

\subsection{SDSS Deep Imaging Data}

The SDSS was an imaging and spectroscopic survey of the sky \citep{yor00} 
using a dedicated wide-field 2.5 m telescope \citep{gun06} at Apache Point
Observatory. Imaging was carried out in drift-scan mode using a 142 mega-pixel
camera \citep{gun98} which gathered data in five broad bands, $ugriz$, 
spanning the range from 3000 to 10,000 \AA\ \citep{fuk96}, on moonless 
photometric \citep{hog01} nights of good seeing. The effective exposure time 
was 54 s. The images were processed using specialized software \citep{lup01}, 
and were photometrically \citep{tuc06,ive04} and astrometrically \citep{pie03} 
calibrated using observations of a set of primary standard stars \citep{smi02} 
on a neighboring 20-inch telescope. All magnitudes are roughly on an AB system 
\citep{oke83}, and use the asinh scale described by \citet{lup99}.

The primary goal of the SDSS imaging survey was to scan 8500 deg$^2$ of the 
north Galactic cap (hereafter referred to as the SDSS main survey). In 
addition to the main survey, SDSS also conducted a deep survey by repeatedly 
imaging a 300 deg$^2$ area on the celestial equator in the south Galactic cap 
in the fall \citep[hereafter referred to as the SDSS deep survey;][]{aba09}. 
This deep stripe (also called Stripe 82; see Stoughton et al.\ 2002) 
spans $\rm 20^h<R.A.<4^h$ and
$\rm -1.25\degr<decl.<1.25\degr$. The multi-epoch images, when coadded, allow
the selection of much fainter quasars than the SDSS main survey.
We used the 10-run co-added data (constructed in 2005) in Paper I and
found five $z\sim6$ quasars with $20<z_{AB}<21$ in this area.

The construction of the co-added images is close, but not identical, to that 
of Annis et al. (in preparation), Paper 1, and \citet{aba09}. We used 
$50\sim60$ scans from the SDSS deep southern stripe, restricting ourselves to 
fields with $r$-band seeing less than $2\arcsec$ and $r$-band sky fainter than 
19.5 mag/arcsec$^2$. Because our selection algorithm uses only $r$, $i$, and
$z$-band photometry, we did not co-add the $u$ and $g$-band data.
The input images are the SDSS corrected frames (or {\it fpC} images).
Each input image was calibrated to a standard SDSS zero point and weighted 
on a field-by-field basis with
\begin{equation}
w=\frac{T}{\rm FWHM^2\,\sigma^2}\,,
\end{equation}
where $T$ is the transparency as measured by the relative zero point of the 
image, FWHM is the full width at half maximum of the PSF, and $\sigma^2$ is 
the variance of the sky. We did not use mask files in the weight map.
After sky background was estimated and subtracted, the images were mapped onto 
a uniform rectangular output astrometric grid using a modified version of the 
registration software SWARP \citep{ber02}. The weight maps were subjected to 
the same mapping. The final co-added images are about two magnitudes deeper 
than the SDSS single-run data. The median seeing of the co-adds as measured in 
the $z$ band is $\sim1.2\arcsec$.

\subsection{Photometry}

The co-added images included in the SDSS DR7 \citep{aba09} were run 
through the SDSS photometric pipeline PHOTO. For a variety of technical 
reasons, we could not do the same for our co-added images, so we used 
SExtractor \citep{ber96} instead of PHOTO to do photometry. For each field, 
the photometry of the three ($riz$) co-added images includes the following 
steps. First we detected sources in the $z$-band image. We used aperture 
photometry with an aperture (diameter) size $3\arcsec$ (or 7.5 pixels), 2.5 
times the typical PSF FWHM. We then used SExtractor double-image mode 
to do photometry in the $r$ and $i$ bands at the positions of the $z$-band 
detections, using the same $3\arcsec$ aperture. Finally we applied aperture 
correction to a large aperture using standard stars in the same field 
\citep{ive07}, and corrected for Galactic extinction using \citet{sch98}.

\section{CANDIDATE SELECTION AND FOLLOW-UP OBSERVATIONS}

\subsection{Quasar Selection Procedure}

Because of the rarity of high-redshift quasars and overwhelming number of
contaminants, our selection procedure of faint $z>5.7$ quasars contains 
several steps. It is similar to the procedure used in Paper I:
\begin{enumerate}
\item Select $i$-dropout sources from the SDSS co-added data.
Objects with $i_{AB}-z_{AB}>2.2$ and $z_{AB}<21.8$ that were not detected in 
the $r$ band were selected as $i$-dropout objects. The color cut
separates high-redshift quasars from the 
majority of stellar objects \citep[e.g.][]{fan01}; $z_{AB}=21.8$ corresponds 
roughly to a 10 $\sigma$ detection, i.e., magnitude errors of $\sim0.1$ mag.
\item Remove false $i$-dropout objects. All $i$-dropout objects were visually
inspected, and false detections were deleted from the list of candidates. 
In Paper I, the majority of the contaminants were cosmic rays. In this paper
we incorporated a sigma clipping algorithm into SWARP during the pixel 
co-addition, which removes almost all cosmic rays as well as high 
proper-motion objects. The remaining cosmic rays were recognized by visually
comparing the individual multi-epoch images making up the co-adds. Roughly
140 objects with $z_{AB}<21.8$ remained at this stage.
\item NIR photometry of $i$-dropout objects. We then carried
out $J$-band photometry of the $i$-dropout objects selected from the previous 
step. The details of the NIR observations are described in Section 3.2 below.
Using the $i-z$ versus $z-J$ color-color diagrams (Figure 2 in Paper I), 
high-redshift quasar candidates were separated from brown dwarfs.
We selected quasars with the following criteria,
\begin{eqnarray}
i-z>2.2\ \ {\rm and} \ \ z-J<0.5\,(i-z)+0.5\,.
\label{selection}
\end{eqnarray}
We also carried out $Y$-band photometry for some candidates, especially those
that were barely 
detected in the $i$ band and thus have large $i-z$ uncertainties. The $Y$ band 
fills in the gap between $z$ and $J$, and the $Y-J$ color efficiently
separates brown dwarfs from high-redshift quasars, since most dwarfs from 
early L to late T have $Y-J$ colors close to 1, while $z\sim6$ quasars usually 
have $Y-J$ colors below 0.8. So we applied the criterion 
\begin{equation}
Y-J<0.8
\end{equation}
to the candidates for which we had $Y$-band photometry. 25 objects remained at 
this stage. Note that the $i$ and $z$ magnitudes are AB magnitudes and the $Y$ 
and $J$ magnitudes are Vega-based magnitudes.
\item Follow-up spectroscopy of quasar candidates. The final step is to carry
out optical spectroscopic observations of quasar candidates to identify
high-redshift quasars. The details of the spectroscopic observations
are described in Section 3.2.
\end{enumerate}

We applied the above selection criteria to the data in the range 
$\rm 310\degr<R.A.<30\degr$. The data contain some ``holes'' in which the 
coadded images were not available. The effective area is $\sim195$ deg$^2$.
In addition to this main search of quasars down to $z_{AB}=21.8$, we also 
selected 35 $i$-dropout objects with $22.0<z_{AB}<22.3$ in the range 
$\rm 0\degr<R.A.<55\degr$. Eight of them passed the criteria of the third 
step. This is to test how deep one can reach with the SDSS co-added images. 

\subsection{NIR Photometry and Optical Spectroscopic Observations}

We obtained $J$ and $Y$-band photometry of the $i$-dropouts 
using the NOAO SQIID infrared camera 
\citep{ell93} on the 4 m Telescope at KPNO and the NIR imager PANIC 
\citep{mar04} on the Magellan telescopes at Chile. The SQIID observations were 
made in 2007 October. SQIID produces simultaneous images of the same field in
the $J$, $H$, and $K$ bands. The pixel size is $0.39\arcsec$ and the field of 
view (FOV) is about $3\arcmin\times3\arcmin$. 
We used a $2\times2$, $2\times3$, or $2\times4$ dither pattern of $15\arcsec$ 
offsets to obtain good sky subtraction and to remove cosmic rays. The exposure 
time at each dither position was 2 min, which was the co-addition of 8 
separate 15 s exposures. The total integration time on individual targets was 
therefore 8, 12, or 16 min. The SQIID data were reduced using the
package `upsqiid' within IRAF\footnote{IRAF
is distributed by the National Optical Astronomy Observatories, which are
operated by the Association of Universities for Research in Astronomy, Inc.,
under cooperative agreement with the National Science Foundation.}.
Briefly, for each object the SQIID data were dark-subtracted and flat-fielded.
The flat field was the median of 30--50 science images taken at the same 
night. The flat field was also used to create a bad pixel mask. After bad 
pixels were repaired by interpolation, 
the sky background was measured and subtracted 
from the science images. Finally the processed science data were combined to
one co-added image. A few standard stars were taken during the night to 
measure the aperture correction and to carry out absolute flux calibration.

PANIC observations in $Y$ and $J$ were made in 2007 October and 2008 
August. The pixel size of PANIC is $0.125\arcsec$ and the FOV is about 
$2\arcmin\times2\arcmin$.
We used a 5-position dither pattern with a dither offset of $10\arcsec$. The 
exposure time at each dither position varied from 60 to 120 s, so the total 
integration time on individual targets was 5--10 min. The PANIC data were 
reduced using the IRAF PANIC package. The basic procedure is similar to what
we did for the SQIID data. After a dark was subtracted and a linearity 
correction was applied, the frames of each object were flat-fielded. The flat 
field was created from twilight images. Then the sky background was measured 
and subtracted. The processed images were corrected for distortion and were 
combined. 

After NIR photometry of the $i$-dropouts, 25 objects with $z_{AB}<21.8$ and 8
objects with $z_{AB}>22$ satisfied the criteria in Section 3.1. Optical 
spectroscopy of these candidates was carried out using the the Red Channel 
spectrograph on the MMT in 2007 November and 2008 October. The observations 
were performed in long-slit mode with a spectral resolution of $\sim10$ \AA.
The exposure time for each target was 15--30 min, which was sufficient to
identify our candidates under normal weather conditions on the MMT. 
If a target was identified as a quasar, several additional exposures were 
taken to improve the spectral quality. 
The quasar data were reduced using standard IRAF routines.

\section{DISCOVERY OF SIX QUASARS AT {\boldmath $z\sim6$}}

From the spectroscopic observations of 32 candidates on the MMT 
we discovered six new quasars at $z\sim6$ in the SDSS deep stripe. 
The other candidates are all late M or L/T dwarfs. 
Figure 1 shows the $z$-band finding charts of the quasars, and Table 1
gives their optical and NIR properties. The $i_{AB}$ and $z_{AB}$ magnitudes 
are taken from the SDSS deep imaging data, and the $Y$ and $J$ magnitudes are 
obtained from our SQIID and PANIC observations. Four of the six quasars were 
discovered in our main quasar search and comprise a flux-limited sample at 
$z_{AB}<21.8$. The other two quasars were found among our $z_{AB}>22$ 
candidates (although they are not a complete sample); their discovery
implies that we can reach deeper than $z_{AB}=22$ in the future.
Figure 2 shows the optical spectra of the six quasars. The total exposure time 
on each quasar was 90--120 min on the MMT. Each spectrum has been scaled to 
the corresponding $z_{AB}$ magnitude given in Table 1, and thereby is on an 
absolute flux scale with an uncertainty of $\sim10$\%.

In Paper I redshifts were measured from either the \lya, \nv\ $\lambda$1240 
(hereafter \nv), or the \oi\ $\lambda1304$ (hereafter \oi) emission line,
but the quasars in this paper are one magnitude fainter, and the spectra do
not have the S/N to detect weak lines. We thus estimate the redshifts from
\lya. Four quasars, SDSS J023930.24--004505.4\footnote{The naming convention 
for SDSS sources is SDSS JHHMMSS.SS$\pm$DDMMSS.S, and the positions are 
expressed in J2000.0 coordinates. We use SDSS JHHMM$\pm$DDMM for brevity.}
(hereafter SDSS J0239--0045),
SDSS J214755.41+010755.3 (hereafter SDSS J2147+0107),
SDSS J230735.35+003149.4 (hereafter SDSS J2307+0031), and
SDSS J235651.58+002333.3 (hereafter SDSS J2356+0023), have prominent \lya\ 
emission lines. For each of them, we measure the \lya\ line center using a 
Gaussian profile to fit the upper $\sim50$\% of the line. Redshifts derived 
from the \lya\ line center are usually biased because the blue side of \lya\ 
is affected by \lya\ forest absorption. The mean shift with respect to the 
systemic redshift at $z>3$ is about 600 km s$^{-1}$ \citep{she07}, 
corresponding to $\delta z\sim0.015$ at $z\sim6$. We correct for this bias for 
the redshifts of the four quasars. The other two quasars, 
SDSS J012958.51--003539.7 (hereafter SDSS J0129--0035) and
SDSS J205321.77+004706.8 (hereafter SDSS J2053+0047), show very weak \lya\ 
emission. Their redshifts are simply estimated from the position of the onset
of sharp \lya\ absorption. The results are listed in Column 2 of Table 1. The 
redshift error of 0.03 quoted in the table is the scatter in the relation 
between \lya\ redshifts and systemic redshifts measured from other lines
\citep{she07}, which is much larger than the statistical uncertainty of our 
fitting process. 

We measure the rest-frame equivalent width (EW) and full width at half maximum
(FWHM) of the \lya\ emission line for each quasar, after fitting and 
subtracting the continuum. The wavelength coverage of each spectrum is too 
short to fit the continuum slope, so we assume that it is a power 
law with a slope $\alpha_{\nu}=-0.5$ ($f_{\nu}\sim\nu^{\alpha_{\nu}}$), and 
normalize it to the spectrum at rest frame 1275--1295 \AA, a continuum window 
with little contribution from line emission. For the four quasars with 
prominent \lya\ emission, we use double Gaussian profiles to fit the broad and
narrow components of \lya. In SDSS J2356+0023, \nv\ is clearly detected and
blended with \lya, so we add an additional Gaussian profile for that line.
Since the blue side of the \lya\ emission line is strongly absorbed by the 
\lya\ forest, we only fit the red side of the line and assume that the 
unabsorbed line is 
symmetric. We ignore the weak \siii\ $\lambda1262$ emission line on the red 
side of \nv. For the two quasars whose \lya\ emission is very weak, we 
calculate the \lya+\nv\ EW by integrating the continuum-subtracted spectra 
over the wavelength range 1216 \AA\ $<\lambda_0<$ 1250 \AA. The measured EW 
and FWHM are shown in Table 2. We also give the FWHM of \lya\ 
in units of km s$^{-1}$. The EW and FWHM of \lya\ for 
SDSS J0239--0045, SDSS J2147+0107, SDSS J2307+0031, and SDSS J2356+0023 have 
taken into account the effect of \lya\ forest absorption; while for
SDSS J0129--0035 and SDSS J2053+0047, the listed EW values include the \nv\ 
line, and were not corrected for \lya\ forest absorption.
The best-fitting power-law continuum is also used to calculate $m_{1450}$ and
$M_{1450}$, the apparent and absolute AB magnitudes of the continuum at
rest-frame 1450 \AA. The results are given in the last two columns of Table 2.

The quasars in this paper have average \lya\ EW and FWHM of 31 \AA\ and 10 
\AA\ (with large scatters of 23 \AA\ and 5 \AA), significantly smaller than 
those in typical low-redshift quasars or more luminous quasars at $z\sim6$
(Paper I).
This is not caused by a selection effect, since quasars with stronger \lya\ 
emission have larger $i-z$ colors and thus are easier to find by our selection 
criteria. The narrowness of the \lya\ emission lines may imply small central 
black hole masses in these high-redshift objects. Black hole masses in quasars 
can be estimated from the widths of broad emission lines; strong UV lines such 
as \civ\ $\lambda$1549 (hereafter \civ) and \mgii\ $\lambda$2800 (hereafter 
\mgii) are frequently used \citep[e.g.][]{mcl04,ves06,she08}. In luminous 
$z\sim6$ quasars black hole masses measured in this way are usually several 
billion solar masses \citep[e.g.][]{jia07,kur07}. Although there is no 
established relation between the width of \lya\ and those of \civ\ and \mgii, 
the two narrow \lya\ line quasars in Paper I (SDSS J000552.34--000655.8 and 
SDSS J030331.40--001912.9) also have narrow \civ\ and \mgii\ lines.
Their estimated black hole masses are only 2-$3\times10^8$ M$_\sun$, and the
corresponding Eddington luminosity ratios are close to 2 \citep{kur07,kur09}.
Therefore, the narrowness of the \lya\ emission lines in this paper could also
indicate small black hole masses and high Eddington luminosity ratios,
suggesting that the black holes in low-luminosity quasars at early epochs 
grow on an Eddington time scale.

\subsection{Notes on individual objects}

{\it SDSS J0129--0035 ($z=5.78$) and SDSS J0239--0045 ($z=5.82$).}
These objects were discovered in our search for quasars with $z_{AB}>22$. They
are by far the faintest $z\sim6$ quasars found by SDSS. SDSS J0129--0035 has a
weak \lya\ emission line; the rest-frame EW of \lya+\nv\ is only 18 \AA.

{\it SDSS J2053+0047 ($z=5.92$).} SDSS J2053+0047 is the brightest quasar in 
this sample. The \lya\ emission line in this quasar is very weak; the 
rest-frame EW of \lya+\nv\ is only 8 \AA. 

{\it SDSS J2147+0107 ($z=5.81$).} SDSS J2147+0107 has the narrowest \lya\
emission line; the line width is only 1500 km s$^{-1}$. If this is typical
of the broad line width in this quasar, the central black hole mass 
would be below $10^8$ M$_\sun$.

{\it SDSS J2307+0031 ($z=5.87$).} SDSS J2307+0031 also has a narrow \lya\
emission line, as we can see that \nv\ is tentatively detected and well 
separated from \lya. The \oi\ emission line also appears to be detected in the
spectrum.

{\it SDSS J2356+0023 ($z=6.00$).} SDSS J2356+0023 is the most distant quasar
in this sample. It has the strongest \lya\ line (EW = 68 \AA). Its \nv\
emission line is also strong. The rest-frame EW and FWHM of \nv\ are 12 \AA\
and 14 \AA, respectively.

\section{QUASAR LUMINOSITY FUNCTION AT $z\sim6$}

We derive the spatial density of the four quasars with $z_{AB}<21.8$ using the 
traditional $1/V_{a}$ method \citep{avn80}. The available volume for a quasar 
with absolute magnitude $M_{1450}$ and redshift $z$ in a magnitude bin
$\Delta M$ and a redshift bin $\Delta z$ is
\begin{equation}
V_{a} = \int_{\Delta M}\int_{\Delta z}p(M_{1450},z) \frac{dV}{dz} dz\,dM,
\end{equation}
where $p(M_{1450},z)$ is the selection function, the probability that a quasar
of a given $M_{1450}$ and $z$ would enter our sample given our selection
criteria. The calculation of the selection function is described in detail in 
Paper I. We use one $M_{1450}$--$z$ bin for our small sample.
The redshift integral is over the redshift range $5.7<z<6.6$ and the magnitude
integral is over the range covered by the sample. The spatial density and its
statistical uncertainty can be written as
\begin{equation}
\rho = \sum_i \frac{1} {V_{a}^{i}}, \ \
\sigma(\rho) = \left[\sum_i \left(\frac{1} {V_{a}^{i}}\right)^2 \right]^{1/2},
\end{equation}
where the sum is over all quasars in the sample. We find that the spatial
density at $\langle M_{1450}\rangle=-25.1$ is $\rho=(7.5\pm3.8)\times10^{-9}$ 
Mpc$^{-3}$ mag$^{-1}$.

In the SDSS main survey, 17 quasars at $z>5.7$ were selected using similar 
criteria and comprise a flux-limited sample with $z_{AB}<20$ over $\sim8000$ 
deg$^2$ (hereafter Sample I). The six quasars of Paper I form a flux-limited 
sample with $z_{AB}<21$ (hereafter Sample II). The QLF at $z\sim6$ based on 
these two SDSS samples and the \citet{coo06} sample with one quasar is well 
fit to a single power law 
$\Phi(L_{1450})\propto L_{1450}^{\beta}$, or,
\begin{equation}
\Phi(M_{1450})=\Phi^{\ast}10^{-0.4(\beta+1)(M_{1450}+26)},
\end{equation}
with $\beta=-3.1\pm0.4$.
In this paper four quasars make a flux-limited sample with $21.0<z_{AB}<21.8$ 
over $\sim195$ deg$^2$ (hereafter Sample III). 
We combine the three SDSS samples to derive the QLF at 
$z\sim6$. The quasars in Sample I are divided into three luminosity bins as 
shown in Figure 3. As described in Paper I, we assume that the bright-end QLF 
is a power law, and we use Equation 5 to fit (i) Samples I and II (the four 
luminous data points in Figure 3) and (ii) Samples I, II, and III (all the data 
points in Figure 3). We only consider luminosity 
dependence and neglect redshift evolution over our narrow redshift range. 
We find that $\beta=-2.9\pm0.4$ and $-2.6\pm0.3$, respectively, and 
the goodness-of-fit in both cases is acceptable ($\chi^2_\nu=0.7$ and 0.9) 
due to large statistical errors.
At low redshift, QLFs can be described by a double power-law characterized
by a break luminosity, and bright-end and faint-end slopes. The current sample 
is not deep enough to reach the break luminosity at $z\sim6$.
The derived slope at $z\sim6$ is slightly flatter than the bright-end slope
at $z\sim2$ but steeper than that at $z\sim4$ \citep{ric06,hop07}.

Sample III does not include any quasars at $z>6.1$. The fraction of $z>6.1$
quasars among the known SDSS quasars at $z>5.7$ is $\sim25$\%, so the 
probability of having no $z>6.1$ quasars among a sample of six is 
$(0.75)^6=0.18$, if they obey the same redshift distribution. This is not a 
small probability, but to test our sensitivity to an as-yet undiagnosed 
selection bias against $z>6.1$ quasars, we recalculated the spatial density of 
Sample III over the redshift range $5.7<z<6.1$. The power-law slope of the QLF 
changes from --2.6 to --2.8, i.e., by less than $1\sigma$.

\section{SUMMARY}

We have discovered six quasars in the redshift range $5.8\le z\le 6.0$ in the 
SDSS deep stripe. The objects are about two
magnitudes fainter than the luminous $z\sim6$ quasars found in the SDSS main 
survey \citep{fan06} and one magnitude fainter than the quasars reported in 
Paper I. The \lya\ emission lines in these quasars are significantly weaker 
and narrower than those in low-redshift quasars or more luminous quasars at 
$z\sim6$. The narrowness of the emission lines may indicate small black hole 
masses and high Eddington luminosity ratios, and therefore short black hole
growth time scales.

Four of the quasars make a flux-limited sample at $z_{AB}<21.8$ over an 
effective area of 195 deg$^2$. The other two quasars were found in a search 
for quasars with $z_{AB}>22$, and do not comprise a complete sample. 
The comoving quasar spatial density at $\langle M_{1450}\rangle=-25.1$ is 
$(7.5\pm3.8)\times10^{-9}$ Mpc$^{-3}$ mag$^{-1}$. We model the QLF at $z\sim6$ 
based on the combination of this sample, the luminous SDSS quasar sample, and 
the Paper I sample. The QLF is well described as a single power law 
$\Phi(L_{1450})\propto L_{1450}^{\beta}$ with a slope $\beta=-2.6\pm0.3$
down to $M_{1450}\sim-25$. The slope changes to $-2.9\pm0.4$ if the new sample
is excluded.

The discovery of the two faintest quasars in this paper indicates that we can 
probe 0.5 magnitude deeper than the complete $z_{AB}<21.8$ sample.
We are constructing new co-added images by
including more available data and by improving our co-addition procedure. 
The new co-added images will be run through PHOTO for accurate PSF photometry.
We hope to obtain a complete sample with $z_{AB}\le22.5$ 
over the full SDSS southern stripe in the next few years.

\acknowledgments

We acknowledge supports from NSF Grants AST-0307384 and AST-0806861 and a 
Packard Fellowship for Science and Engineering (LJ, XF, and FB).
XF also acknowledges support from Max Planck Society.
MAS acknowledges the support of NSF Grant AST-0707266.
We would like to thank the MMT staff, Magellan staff, and KPNO staff for their 
expert help in preparing and carrying out the observations.

Funding for the SDSS and SDSS-II has been provided by the Alfred P. Sloan 
Foundation, the Participating Institutions, the National Science Foundation, 
the U.S. Department of Energy, the National Aeronautics and Space 
Administration, the Japanese Monbukagakusho, the Max Planck Society, and 
the Higher Education Funding Council for England. The SDSS Web Site is 
http://www.sdss.org/.
The SDSS is managed by the Astrophysical Research Consortium for the 
Participating Institutions. The Participating Institutions are the American 
Museum of Natural History, Astrophysical Institute Potsdam, University of 
Basel, University of Cambridge, Case Western Reserve University, University 
of Chicago, Drexel University, Fermilab, the Institute for Advanced Study, 
the Japan Participation Group, Johns Hopkins University, the Joint Institute 
for Nuclear Astrophysics, the Kavli Institute for Particle Astrophysics and 
Cosmology, the Korean Scientist Group, the Chinese Academy of Sciences 
(LAMOST), Los Alamos National Laboratory, the Max-Planck-Institute for 
Astronomy (MPIA), the Max-Planck-Institute for Astrophysics (MPA), 
New Mexico State University, Ohio State University, University of Pittsburgh, 
University of Portsmouth, Princeton University, the United States Naval 
Observatory, and the University of Washington.

{\it Facilities:} 
\facility{Sloan}, 
\facility{Mayall (SQIID)}, 
\facility{Magellan:Baade (PANIC)}, 
\facility{MMT (Red Channel spectrograph)}

\clearpage
\begin{deluxetable}{cccccc}
\rotate
\tablecaption{Optical and NIR Photometry}
\tablewidth{0pt}
\tablehead{\colhead{Quasar (SDSS)} & \colhead{Redshift\tablenotemark{a}} &
  \colhead{$i_{\rm AB}$ (mag)} &  \colhead{$z_{\rm AB}$ (mag)} &
  \colhead{$Y$ (mag)} & \colhead{$J$ (mag)} }
\startdata
J012958.51--003539.7 & 5.78$\pm$0.03 & 24.52$\pm$0.25 & 22.16$\pm$0.11 & $\ldots$       & 21.78$\pm$0.15 \\
J023930.24--004505.4 & 5.82$\pm$0.03 & 25.40$\pm$0.60 & 22.08$\pm$0.11 & 21.62$\pm$0.05 & 21.15$\pm$0.11 \\
J205321.77+004706.8  & 5.92$\pm$0.03 & 24.35$\pm$0.29 & 21.41$\pm$0.06 & $\ldots$       & 20.46$\pm$0.07 \\
J214755.41+010755.3  & 5.81$\pm$0.03 & 24.00$\pm$0.21 & 21.61$\pm$0.08 & 20.92$\pm$0.18 & 20.79$\pm$0.14 \\
J230735.35+003149.4  & 5.87$\pm$0.03 & 26.62$\pm$2.11 & 21.77$\pm$0.09 & 20.99$\pm$0.16 & 20.43$\pm$0.11 \\
J235651.58+002333.3  & 6.00$\pm$0.03 & 24.52$\pm$0.25 & 21.66$\pm$0.08 & $\ldots$       & 21.18$\pm$0.07 \\
\enddata
\tablenotetext{a}{The redshift error of 0.03 is the scatter in the relation 
between \lya\ redshifts and systemic redshifts measured from other lines 
\citep{she07}, which is much larger than the statistical uncertainties 
measured from our fitting process.}
\tablecomments{The $i_{\rm AB}$ and $z_{\rm AB}$ magnitudes are AB magnitudes 
   and the $Y$ and $J$ magnitudes are Vega-based magnitudes.}
\end{deluxetable}

\clearpage
\begin{deluxetable}{cccccc}
\rotate
\tablecaption{Properties of Continua and Emission Lines}
\tablewidth{0pt}
\tablehead{\colhead{Quasar (SDSS)} & \colhead{Redshift} & \colhead{EW (\lya)} &
\colhead{FWHM (\lya)} & \colhead{$m_{1450}$ (mag)} & \colhead{$M_{1450}$ (mag)}}
\startdata
J0129--0035 & 5.78 & 18$\pm$3 & $\ldots$                         & 22.28$\pm$0.12 & --24.36$\pm$0.12 \\
J0239--0045 & 5.82 & 46$\pm$8 & 15$\pm$3 (3700 $\rm km\,s^{-1}$) & 22.15$\pm$0.12 & --24.50$\pm$0.12 \\
J2053+0047  & 5.92 &  8$\pm$1 & $\ldots$                         & 21.20$\pm$0.07 & --25.47$\pm$0.07 \\
J2147+0107  & 5.81 & 28$\pm$3 &  6$\pm$2 (1480 $\rm km\,s^{-1}$) & 21.65$\pm$0.10 & --25.00$\pm$0.10 \\
J2307+0031  & 5.87 & 15$\pm$4 &  7$\pm$2 (1730 $\rm km\,s^{-1}$) & 21.73$\pm$0.10 & --24.93$\pm$0.10 \\
J2356+0023  & 6.00 & 68$\pm$5 & 13$\pm$2 (3200 $\rm km\,s^{-1}$) & 21.77$\pm$0.10 & --24.92$\pm$0.10 \\
\enddata
\tablecomments{Rest-frame FWHM and EW are in units of \AA. For 
SDSS J0239--0045, SDSS J2147+0107, SDSS J2307+0031, and SDSS J2356+0023, the 
\lya\ EW and FWHM have taken into account absorption by the \lya\ 
forest; while for SDSS J0129--0035 and SDSS J2053+0047, the listed EW values 
include the \nv\ line, and were not corrected for \lya\ forest absorption.}
\end{deluxetable}

\clearpage
\begin{figure}
\epsscale{1.0}
\plotone{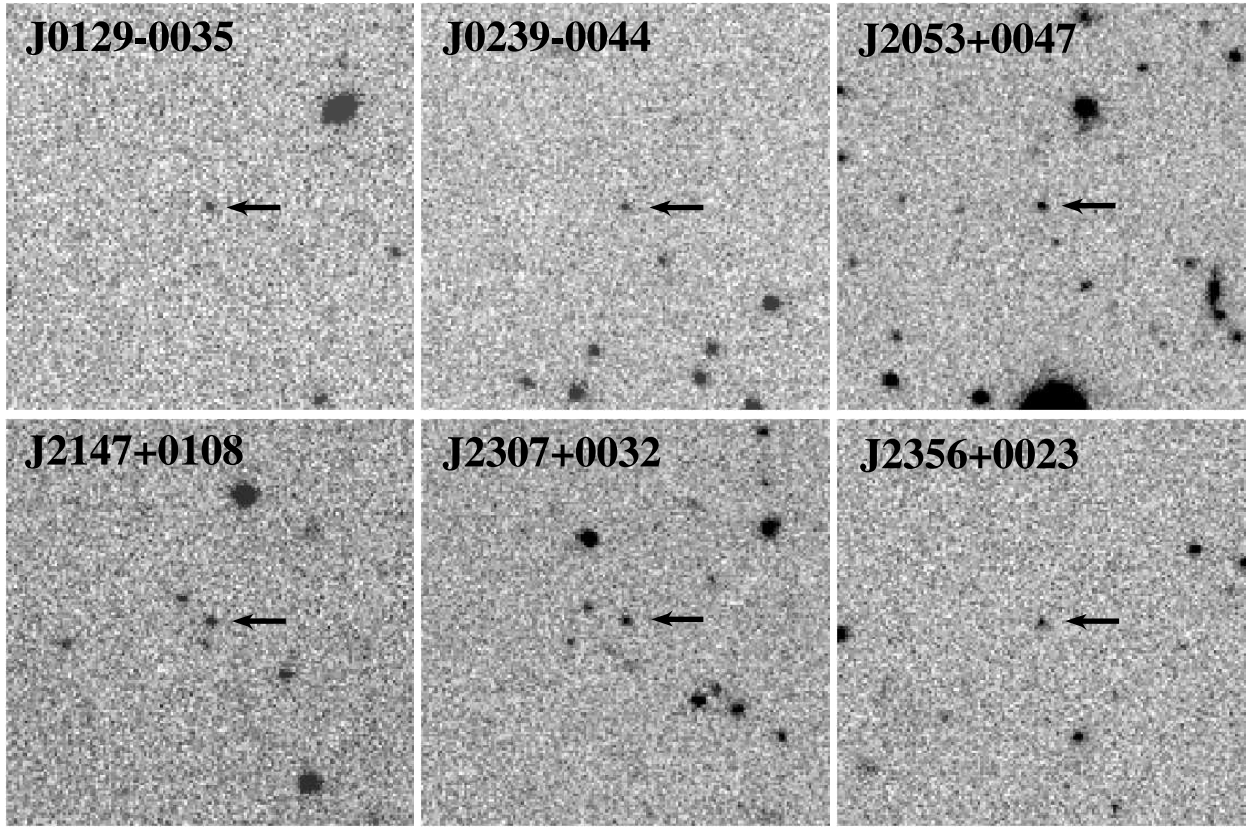}
\caption{The $z$-band finding charts of the six new $z\sim6$ quasars
discovered in the SDSS deep stripe. The charts are from the co-added images 
with 50--60 SDSS runs. The size is $1'\times1'$.
North is up, and east to the left.}
\end{figure}

\clearpage
\begin{figure}
\epsscale{0.8}
\plotone{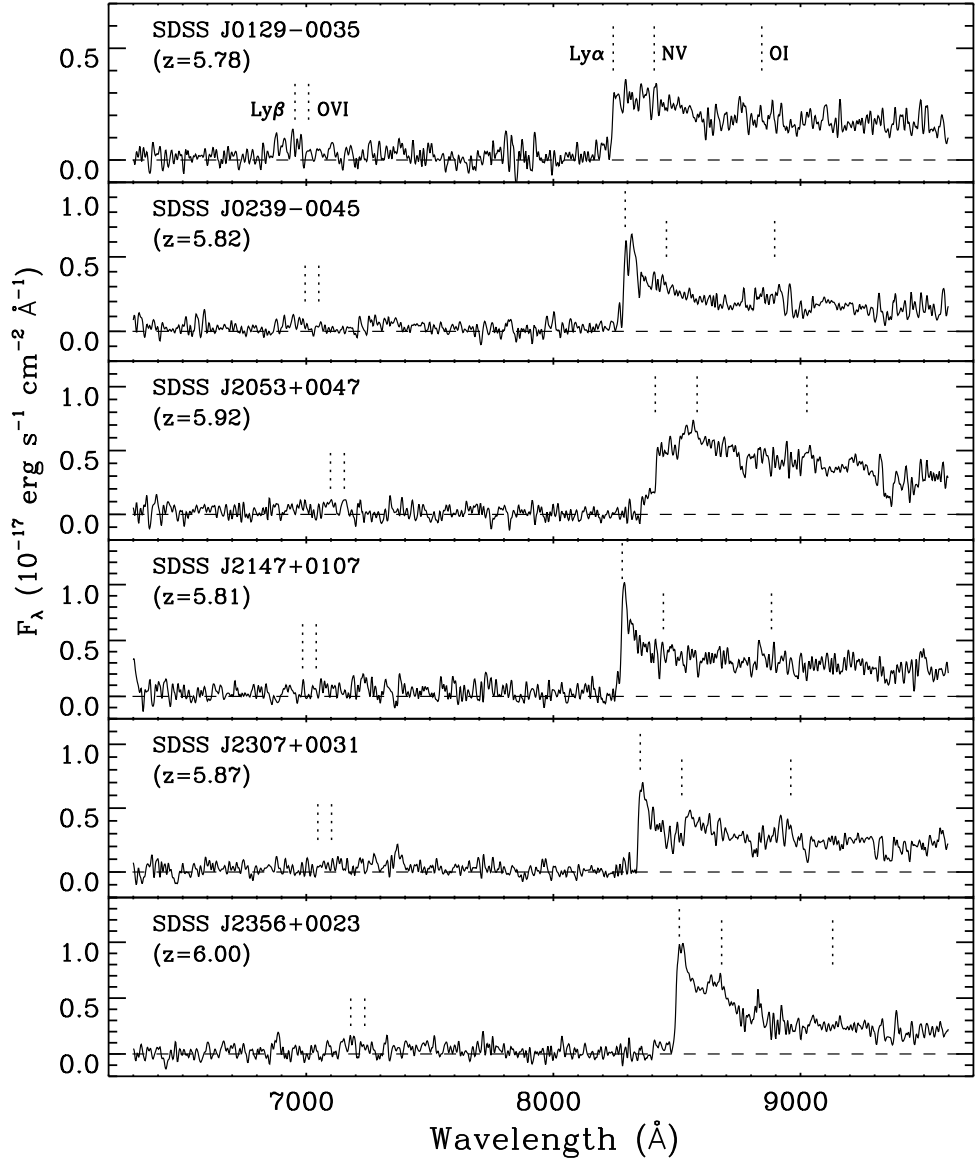}
\caption{Optical spectra of the six new $z\sim6$ quasars discovered in the
SDSS deep stripe. The spectra were taken on the MMT Red Channel 
with a spectral resolution of $\sim10$ \AA.
The total exposure time on each quasar was 
90--120 min. The spectra have been smoothed by three pixels and scaled 
to the corresponding $z_{AB}$ magnitudes given in Table 1.}
\end{figure}

\clearpage
\begin{figure}
\epsscale{0.8}
\plotone{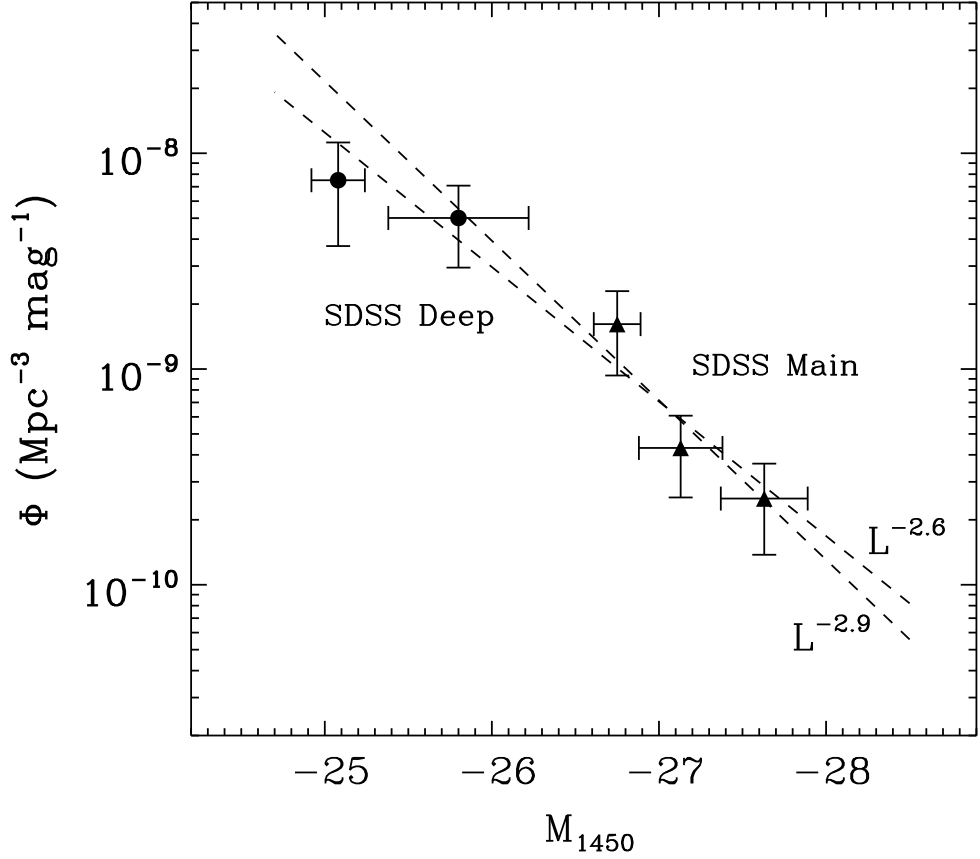}
\caption{The quasar luminosity function at $z\sim6$. The horizontal axis is 
the absolute AB magnitude of the continuum at rest-frame 1450 \AA.
The filled circles represent the $1/V_{a}$ densities 
of the quasars discovered in the SDSS deep stripe, and the filled triangles 
represent the densities from a study of 17 quasars from the SDSS main survey.
The dashed and dotted lines show the best power-law fits to the four luminous 
data points (the luminous SDSS sample and the Paper I sample) and all the data 
points (all the three samples), respectively. The goodness-of-fit is 
acceptable in both cases ($\chi^2_\nu=0.7$ and 0.9).}
\end{figure}

\end{document}